\documentstyle[12pt,epsf]{article}
\newcommand{\rf}[1]{(\ref{#1})}
\newcommand{\bea}{\begin{eqnarray}}
\newcommand{\eea}{\end{eqnarray}}

\renewcommand{\l}{\lambda}

\newcommand{\n}{\nu}
\newcommand{\m}{\mu}

\newcommand{\ra}{\right\rangle}
\newcommand{\la}{\left\langle}

\newcommand{\cD}{{\cal D}}

\newcommand{\cM}{{\cal M}}

\def\void{}
\def\labelmark{}

\newenvironment{formula}[1]{\def\labelname{#1}
\ifx\void\labelname\def\junk{\begin{displaymath}}
\else\def\junk{\begin{equation}\label{\labelname}}\fi\junk}%
{\ifx\void\labelname\def\junk{\end{displaymath}}
\else\def\junk{\end{equation}}\fi\junk\labelmark\def\labelname{}}

{\ifx\void\labelname\def\junk{\end{array}\end{displaymath}}
\else\def\junk{\end{array}\right.\end{equation}}
\fi\junk\labelmark\def\labelname{}\def\junk{}
}

\newcommand{\beq}{\begin{formula}}
\newcommand{\eeq}{\end{formula}}
\newcommand{\beqv}{\begin{formula}{}}

\begin{document}
\topmargin 0pt
\oddsidemargin 5mm
\headheight 0pt
\headsep 0pt
\topskip 9mm

\hfill    NBI-HE-94-49

\hfill October 1994

\begin{center}
\vspace{24pt}
{\large \bf Computational ergodicity of $S^4$}

\vspace{24pt}

{\sl J. Ambj\o rn } and {\sl J. Jurkiewicz}\footnote{Permanent address:
Inst. of Phys., Jagellonian University.,
ul. Reymonta 4, PL-30 059, Krak\'{o}w~16, Poland}

\vspace{6pt}

 The Niels Bohr Institute\\
Blegdamsvej 17, DK-2100 Copenhagen \O , Denmark\\

\end{center}
\vspace{24pt}

\addtolength{\baselineskip}{0.20\baselineskip}
\vfill

\begin{center}
{\bf Abstract}
\end{center}

\vspace{12pt}

\noindent
It is known that there are four-manifolds which are not
algorithmically recognizable. This implies that there
exist triangulations of these manifolds which are separated by
large barriers from the point of view of the computer algorithm.
We have not observed these barriers for triangulations of $S^4$.

\vfill

\newpage

\section{Introduction}

A non-perturbative formulation of quantum gravity is one of the greatest
challenges of theoretical physics today. One such suggestion which
has attracted certain attention in the last couple of years is to
use so-called ``dynamical triangulations''\cite{aj,am}. This approach
provides us with a regularization of Euclidean quantum gravity, much in the
spirit of the lattice regularization of Euclidean quantum field theories
where the continuum limit is recovered at points in coupling
constant space where the statistical systems have second order transitions.
One difference is due to the dynamical nature of space-time
in a theory of gravitation. This forces us to use ``dynamical triangulations''
of space-time rather than a fixed lattice: We have to sum over classes
of different triangulations in the path integral. The partition function
of Euclidean quantum gravity for a compact, closed manifold $\cM$
can be written as:
\beq{*1}
Z = \int \frac{\cD g_{\m\n}}{{\rm Vol} (diff)} \; e^{-S[g]},
\eeq
where ${\rm Vol} (diff)$ is the ``volume'' of the diffeomorphism group
of $\cM$ and the integration is over
equivalence classes of Riemannian structures on $\cM$.
It is no loss of generality to view $\cM$ as a combinatorial or, equivalently,
piecewise linear manifold, since there is a one-to-one correspondence
between smooth and piecewise linear structures for manifolds of dimensions
$D$ less than seven. $S[g]$ is the gravitational action, which we here
will take to be the Einstein-Hilbert action:
\beq{*1a}
S[g] = \l \int d^D \tau \; \sqrt{g} - \frac{1}{16\pi G} \int d^D \tau
\; \sqrt{g} R.
\eeq
The regularized version of this functional integral  in the context
of dynamical triangulations is replaced by
\beq{*2}
Z = \sum_{T} \frac{1}{C_T}e^{-S_T}
\eeq
where the summation is over all triangulations of
the manifold $\cM$. To be more precise we consider two
triangulations to be identical if there exists a mapping of the
vertices of one triangulation on  the vertices of the other
triangulation, such that all simplexes (and sub-simplexes) of the
first triangulation are mapped onto the corresponding simplexes
(and sub-simplexes) of the second triangulation. The number $C_T$ is
the order of the automorphism group of the triangulation.
This way of identifying triangulations is compatible with
the introduction of a distance function on the  triangulations
by assigning a lattice length {\it a} to all links and considering the
triangulations as piecewise linear manifolds. In the following we always
take {\it a=1}, but the continuum limit should
always refer to distances large compared to {\it a}, which in this way
serves as a cut off.  By this length assignment
for each $T$ there will be an associated metric assigned to $\cM$,
and by considering all triangulations of $\cM$
we get a grid in space of metrics on $\cM$.
For a given triangulation $T$ the discretized action $S_T$ is evaluated
by Regge calculus and after
some trivial algebra the Regge version of \rf{*1a} for a given
triangulation $T$ with the above distance function reads:
\beq{*3}
S_T = k_4 N_4(T) -k_2 N_2(T),
\eeq
where $N_4(T)$ denotes the number of 4-simplexes and $N_2(T)$ the number
of 2-sim\-plexes in the triangulation $T$. The coupling constant $1/k_2$
is proportional to the bare Einstein coupling constant
$G$ in \rf{*1a}, while $k_4$ is  related to the bare
cosmological constant $\l$ in \rf{*1a}.

If we consider two-dimensional Euclidean quantum gravity we can solve
the continuum theory by conformal field theory methods.
In addition the regularized version
can be solved explicitly and at the critical point of the
statistical model we recover the continuum results obtained
by conformal field theory. When we move to higher dimensions we enter
unchartered territory from the point of view of the continuum version
of Euclidean quantum gravity. Nevertheless \rf{*2}-\rf{*3} provide us
with a perfectly well defined statistical
model and we can search for critical points where
one can attempt to define a non-perturbative continuum limit.

Until now it has only been possible to analyze the model given by
\rf{*2} and \rf{*3} by numerical methods, more specifically by Monte
Carlo simulations\cite{ajk,bm,ckr}. In order to get around in the class of all
triangulations of a combinatorial manifold $\cM$ one needs a set of moves
which can be implemented on the computer and which are ergodic
in the set of triangulations of $\cM$.
Two triangulations of $\cM$ are {\it combinatorially equivalent}, i.e.
they have a common subdivision (up to relabelling of vertices).
It is by now well known that for manifolds of
dimensions $D \leq 4$ there exists a finite set of ``local'' moves
which are ergodic the in class of combinatorially
equivalent triangulations of a
given manifold \cite{gv}. By ``local'' we mean that
each of the moves will change the triangulations
only in such a way that the number of integer algorithmic
operations needed to implement a move is bounded by a fixed
number independent of the triangulation.

Until now the setup seems perfect from a computational point
of view: We have a set of local moves which can connect any
two triangulations of $\cM$ in a finite number of steps. We
have an action, and by Monte Carlo simulations we should now be
able to make an importance sampling of the triangulations with
the weight provided by $e^{-S_T}$, $S_T$ given by \rf{*3}.
However, the fact that there exist four-manifolds
which are algorithmically unrecognizable casts some doubts on this program.
Denote such a manifold by $\cM_0$ and let it be finitely presented
by a combinatorial triangulation $T(\cM_0)$. The algorithmic
unrecognizability of $\cM_0$ means that there exists no algorithm
which allows us to decide whether another manifold $\cM$, again
finitely presented by a triangulation $T(\cM)$ is
combinatorially equivalent to $\cM_0$. When this is combined with
the existence of the finite set of local moves which are able to
connect any two triangulations of $\cM_0$ in a finite
number of steps, but where this number is a function of the
chosen triangulations, one can prove the following
theorem \cite{benav}:\\
{\it The number of moves needed to connect two triangulations of $\cM_0$,
$T$ and $T'$ with $N_4(T)=N_4(T')$, cannot be bounded by any recursive
function $r(N_4)$.} \\
Recall that $\exp(N)$ or
$\exp(\exp(\ldots \exp(N)))$ (the exponentiation $N$ times) are recursive
functions. Effectively this implies that there will be very large
barriers between some classes of triangulations of $\cM_0$ and
there would be triangulations which could never be reached in any reasonable
number of steps even
for quite moderate values of $N_4$. Of course the number of
configurations which are separated from some standard triangulation of
$\cM_0$ by  such barriers could vanish relative to the total number
of configurations as a function of $N_4$. In ref. \cite{benav}
it was conjectured that it will not be the case and some
plausibility arguments in favor of the conjecture were given.
If the conjecture is correct, a Monte Carlo method
based on the finite set of local moves will never get around {\it
effectively} in the class of triangulations of $\cM_0$.
We can say that the moves, although ergodic in the class
of triangulations, will not be {\it computationally ergodic} \cite{benav}.

How is the situation for $S^4$ which is the manifold which
until now has been used in the Monte Carlo simulations ?
It is unknown whether $S^4$ is algorithmically recognizable
in the class of four-manifolds. If $S^4$ is algorithmically
unrecognizable the arguments given above for $\cM_0$ apply.
Since the number of different triangulations with a fixed
$N_4$ is bounded by some number $N_4^n!$, and for triangulations
of a fixed topology, like $S^4$, probably even exponentially bounded,
the only way the number of steps needed to connect any two triangulations
with $N_4$ simplexes can end up not being bounded by a recursive function
is the following: In the process of connecting  two triangulations
by a sequence
of moves we will be forced to very high values of $N_4$. In fact this number
itself cannot be bounded by a recursive function.

It would be unnatural if this phenomenon was not present at essentially all
scales and one would therefore expect to be able to observe it in the
following way: Let us by Monte Carlo simulations generate a number of
independent configurations for some large values of $N_4$.
Now ``shrink'' (again by Monte Carlo simulations) these configurations
to the minimal triangulation of $S^4$, consisting of 6 4-simplexes.
In case we never get seriously stuck in this shrinking procedure
there can be no barrier separating two triangulations since
we can first move to the minimal configuration and then out to another
triangulation by the reverse set of moves.

We have never observed that the triangulations get stuck in the process
of a ``reasonable shrinking procedure''.
We take this as some evidence in favor of $S^4$ being
computationally recognizable.

In the rest of this paper we explain in what sense we are scanning the space
of triangulations of $S^4$ by Monte Carlo simulations, what we mean by
``shrinking'' and we give a tentative ``experimental'' upper limit
of the number of moves needed to connect any two triangulations of $S^4$.
Finally we discuss some implications for quantum gravity.

\section{Scanning the configuration space}

A four-dimensional triangulation is characterized by the number of
vertices, links, triangles, tetrahedra and 4-simplexes which constitute
the triangulation, and most importantly, the information about the
way they are glued together to form a combinatorial four-manifold.
This last requirement means that the numbers $N_n$ of $n$-simplexes,
$n \leq 4$, can not be chosen arbitrary: They have to satisfy the
Dehn-Sommerville relations:
\beq{*5}
N_n = \sum_{i = n}^4 \frac{(-1)^{i+1}(i+1)!}{(n+1)!(i-n)!}N_i.
\eeq
These relations are valid in dimensions $D$ other than four if we replace
$4$ with $D$. They express the manifold requirement that the set of
$D$-simplexes having a $n$-simplex
in common should constitute a  combinatorial $D-n$-ball. If we define
$N_{-1} \equiv \chi$, the Euler characteristics of the manifold,
\rf{*5} gives an additional constraint when $D$ is even. For $D=4$ this
implies that at most two of the $N_n$'s can be chosen independently.
Let it be $N_4$ and $N_2$. It is now truly remarkable that these
are precisely the quantities which enter in the Einstein-Hilbert action
\rf{*3}. The interpretation of the Monte Carlo simulations where we
include the gravitational action is now that the choice of coupling
constants $k_4$ and $k_2$ will determine the average value of the $N_n$'s
but that all triangulations with the same values of $N_n$ will be
chosen with the same probability. By monitoring $k_2$ and $k_4$ we
can explore the neighborhood of the class of triangulations characterized
by a given allowed choice of $N_n$'s. If we increase the average volume
$\la N_4 \ra$ the other $\la N_n \ra$, $n=0,1,2,3$ will increase too.
While from \rf{*5} it follows that there exist positive constants
$a_n,b_n$ such that
$a_n N_4 \leq N_n \leq b_n N_4$ for $n >0$, the number of vertices, $N_0$
behaves differently, and it is easy to construct explicitly triangulations of
$S^4$ such that $N_0 \sim N_4^{1/2}$ as well as
triangulations where $N_0 \sim N_4$ for $N_4 \to \infty$. The first
kind of triangulations seems a little pathological from the point
of view of smooth manifolds since the number of vertices per unit volume
goes to zero. Nevertheless they might be numerous and cannot be dismissed
{\it a priori} in a theory of quantum gravity.

It is easy to understand that triangulations with as few vertices
as possible are favored in the limit $k_2 \to -\infty$ while triangulations
with a maximal number of vertices are favored for $k_2 \to \infty$.
If we use \rf{*5} we can write the Einstein-Hilbert action \rf{*3} as:
\beq{*6}
S_T = k_4' N_4 -2 k_2(N_0-2),~~~~~~~k'_4= k_4-2k_2,
\eeq
and it is seen that $k_2$ acts as a ``chemical'' potential not only for
the number of triangles but also for the number of vertices $N_0$.
In addition it is easy to understand the two limits in a qualitative
way. If there are very few vertices relative to the number of $N_n$'s,
$n >0$, the order of some of the vertices must be quite  high and most likely
it will be possible to move between any two vertices along the links
in few steps. This indicates that the Haussdorff dimension of the
triangulation might be high, maybe even infinite. On the other hand
we get a maximum number of vertices relative to the number of
four-simplexes if we glue four-simplexes together in an almost
one-dimensional structure. These qualitative aspects of the
triangulations as functions of the bare gravitational coupling
constant $1/k_2$ are clearly seen in the Monte Carlo simulations, and
from this point of view the simulations certainly pick
up ``typical'' configurations.

Intuitively it seems as if large barriers may appear
more easily if the manifold is highly crumpled and of
large Hausdoff dimension. {\it A posteriori} these were indeed
the manifolds  which it took the longest time to ``cool''
to the minimal volume configuration of $S^4$. We therefore
concentrated on simulations with small values of $k_2$.
For $k_2=0$ we are well into the region of small $k_2$ from
a practical point of view and simulations performed there
will generate ``typical'' crumpled configurations. The choice $k_2=0$
has the additional nice feature that it weights all triangulations
equally. We performed the main series of
numerical experiments on thermalized configurations with $k_2=0$ and
$N_4=$16000, 32000 and 64000. These configurations were obtained
in numerical experiments, where the system was forced to stay in the
neighborhood of a chosen volume $N_4^0$ by modifying the action \rf{*3}
to:
\beq{*7}
S_T = k_4 N_4(T) + \Delta |N_4-N_4^0| -k_2 N_2(T),
\eeq
with small $\Delta$ and $k_4$ close to the pseudo--critical value
$k_4^c(k_2,N_4)$. The dependence on $N_4$ is a finite--size effect. For
$k_2=0$ the $k_4^c$ values for $N_4=16000,$ 32000 and 64000 were found
to be respectively 1.134(2), 1.152(2) and 1.168(2). The numerical
simulation means performing ``moves''. A number of successfully performed
moves can be used as a measure of time or numerical distance covered
by the simulation. This number for thermalized configurations is typically
of the order $10^9$.

The objective of the experiment was to reduce the volume $N_4$ in the
``cooling'' experiment. One can imagine many possible setups for such
experiment. As is clear from the discussion above small values of $N_4$
will be favored for large $k_4$, provided $k_4 > k_4^c(k_2,N_4)$.
If we choose this value to be very large, the system can be frozen into
one of the metastable configurations and we can try to measure the height
of the barrier separating it from the path leading to the bottom of
the configuration space -- a configuration with a minimal four--volume.
In all experiments presented here we chose $k_2=0$, but the same qualitative
behaviour is seen for other values of $k_2$.
The outcome of a series of typical experiments is presented on figure 1.
In the first step of the experiment the configurations were cooled with
$k_4=8$. The unit of ``time'' on the horizontal axis is 5000 accepted
moves. In few time steps (typically below 10) the system reaches a stable
volume, where only ``canonical'' moves can be performed. Further reduction
of the volume becomes impossible, because the number of points with
coordination 5 and links with coordination 4, necessary to perform the
volume--reducing moves becomes zero and the inverse moves are exponentially
suppressed. The situation can be viewed as reaching the boundary of the
configuration space. In the second step of the experiment the value of
$k_4$ is raised to 6.0. This is still far above the critical value, which
is of the order 1. It is nevertheless sufficient for the system to find
it's way down, eventually reaching the minimal configuration. The height
of the barrier is finite, the system has to increase the volume only
by few (typically 2 -- 8) simplexes before it can be reduced again.

A picture which emerges from this experiment is that of a very smooth
configuration space boundary rather than that with many very deep valleys.
The volume falls down almost linearly with ``time''. For the number
of vertices this dependence is more complicated: the decrease gets faster
for smaller volumes. The corresponding plots are shown on figure 2.

We tried to modify in various ways the first step of this experiment to
get different starting points. In all cases the qualitative behaviour was
the same, although we observed some dependence of the time necessary for
a complete cooling of the initially frozen configuration.

On figure 3 we show results of a series of experiments for the system
with 32000 simplices. In all experiments the same
thermalized $k_2=0$ configuration was cooled with different values of
$k_4 > k_4^c$. We started with $k_4=2.0$, where we observe a smooth volume
dependence; the system never reaches the boundary of the configuration space
and reaches the minimal configuration after 135 steps.
For $k_4 \ge 3.0$ the cooling process has two phases. In the first one
(approximately 10 time steps, independent of $k_4$) the system reaches
the boundary with no vertices with coordination 5 and links with coordination
4.
In the second phase the system slides down, eventually to reach the minimal
configuration. The structures, necessary to perform the volume--reducing
moves are dynamically created (typically one link with coordination 4 and more
rarely a vertex with coordination 5) which is enough to find a path down.
The process gets more difficult the bigger is the value of $k_4$ when the
volume--increasing moves become more suppressed. Figure 4 shows the
corresponding dependence of the number of vertices.

We repeated the experiment for other thermalized configurations
for systems with 32000 simplexes. In the cooling process we set $k_2=0.0$.
As expected,
for configurations typical for larger values of $k_2$ the complete
cooling of the configuration is achieved faster. Already for the
$k_2=1.0$ configuration the cooling time with $k_4=6.0$ takes less than 20
steps.
In the other extreme, we studied configurations for negative $k_2$.
The cooling of the thermalized configuration for $k_2=-1.$
looks almost identical to
that of the $k_2=0.0$ configuration. It should be noticed that for the
$k_2=-1.0$
the number of points $N_0 \approx 500$, so we are dealing with an
extremely crumpled manifold with many points of very high order.

\section{Discussion}

We have not seen any trace of the very large distances, measured in the
number of local moves, which separate  certain configurations
in algorithmically unrecognizable manifolds. This indicates that
$S^4$ is either algorithmically recognizable in the class of
four-manifolds, or that the class of configurations separated
from the trivial minimum configuration is small, maybe
of measure zero, in the class of all configurations of $S^4$.

Since we here talk about numerical ``experiments'' the above results can
not constitute a proof in any way. However we find it remarkable
that we have not seen any sign at all of even small barriers separating
parts of the configuration space from the trivial minimum configuration.
Rather, it seems as if the number of moves needed to connect any two
configurations of  volume $N_4$ is simply proportional to $N_4$.

\vspace{12pt}

\noindent
{\bf Acknowledgement} It is a pleasure to thank Radi Ben-Av for explaining
to us the implications of his theorem.

\addtolength{\baselineskip}{-0.20\baselineskip}

\input rotate

\newpage

\begin{figure}
\vbox{\hbox{\epsfxsize=15.cm\epsfbox{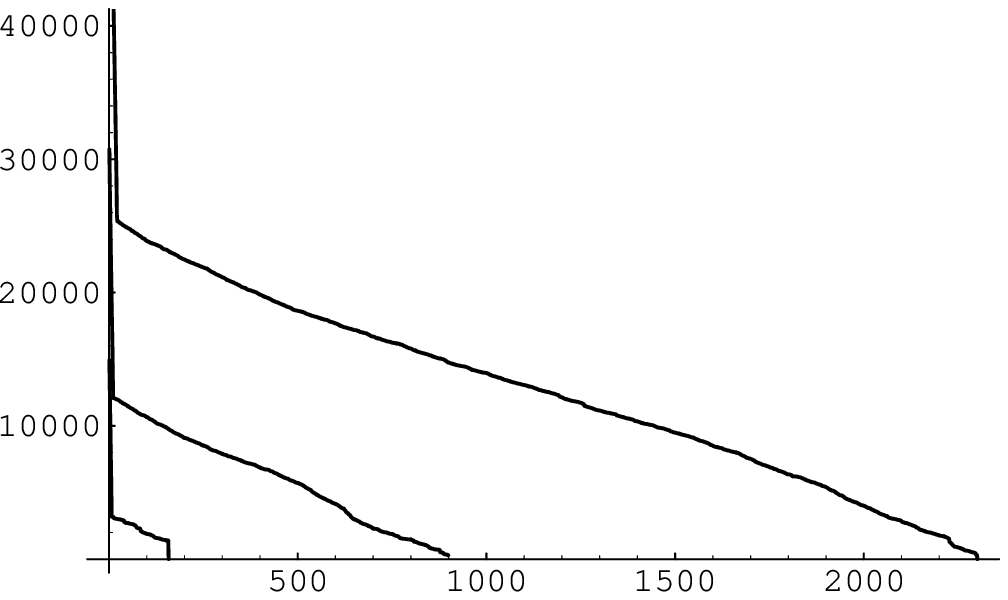}}}
\end{figure}
\noindent
{\bf Fig.1} \\
$N_4$ {\em vs.} time dependence in the
cooling experiments for systems with 16000, 32000 and 64000 simplices. In the
first step systems were cooled with $(\kappa_2=0.0$,
$\kappa_4=8.0)$ to reach a stable pseudo--minimum.
The rest of the cooling was done with $(\kappa_2=0.0$, $\kappa_4=6.0)$
Unit of time is 5000 moves.\\
\newpage
\begin{figure}
\vbox{\hbox{\epsfxsize=15.cm\epsfbox{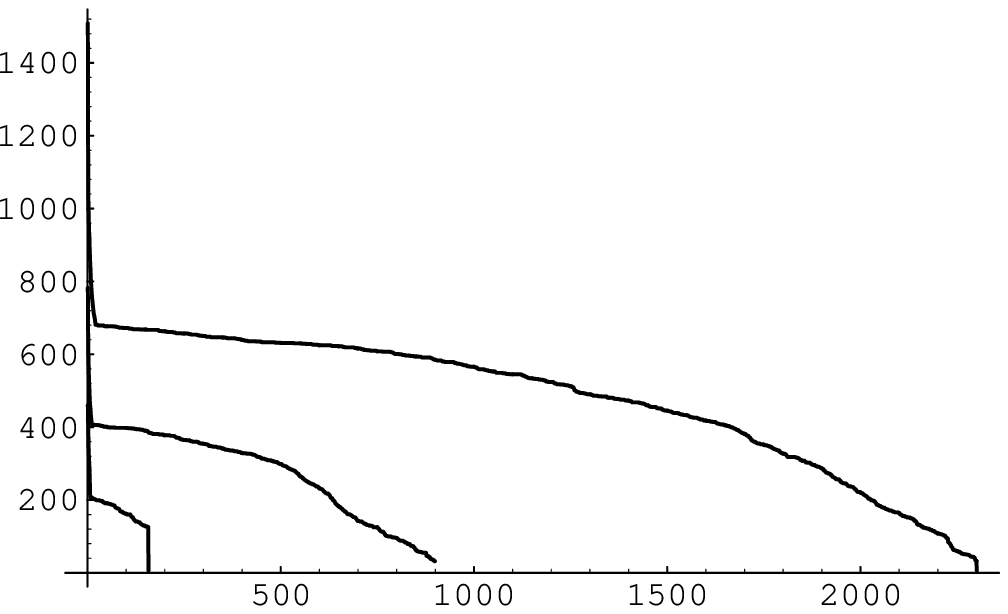}}}
\end{figure}
\noindent
{\bf Fig.2} \\
The same as figure 1, dependence $N_0$ {\em vs.} time.\\
\newpage
\begin{figure}
\vbox{\hbox{\epsfxsize=15.cm\epsfbox{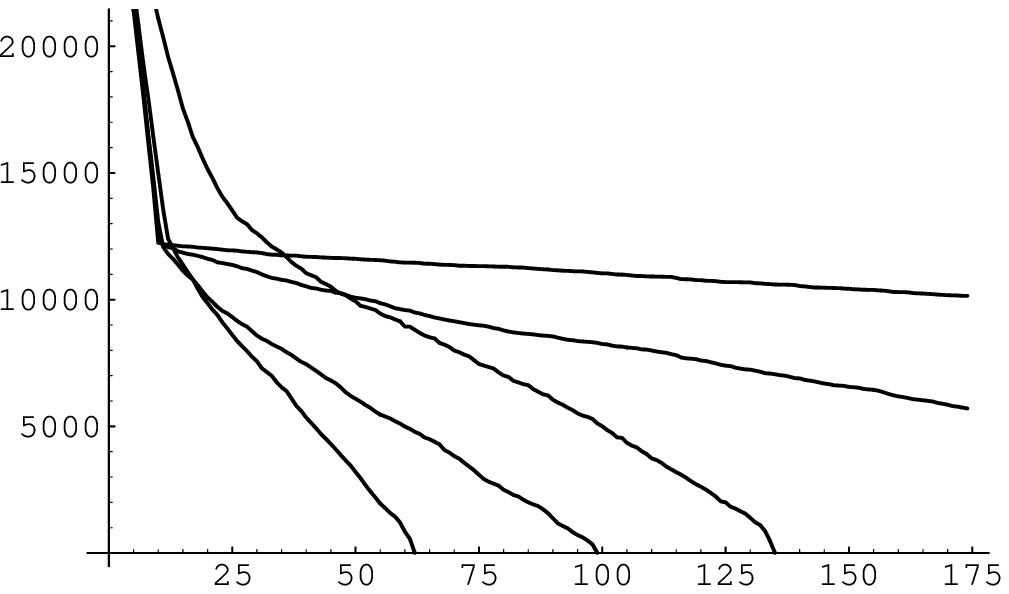}}}
\end{figure}
\noindent
{\bf Fig.3} \\
$N_4$ {\em vs.} time dependence in the
cooling experiments for  a system with 32000 simplices. In all
cases $\kappa_2=0.0$. Plots correspond to $\kappa_4=2.0,3.0,4.0,5.0$ and 6.0
Above $\kappa_4=3.0$ system reaches the ''boundary'' in the first few steps.
In all cases the minmal configuration was reached.\\

\newpage
\begin{figure}
\vbox{\hbox{\epsfxsize=15.cm\epsfbox{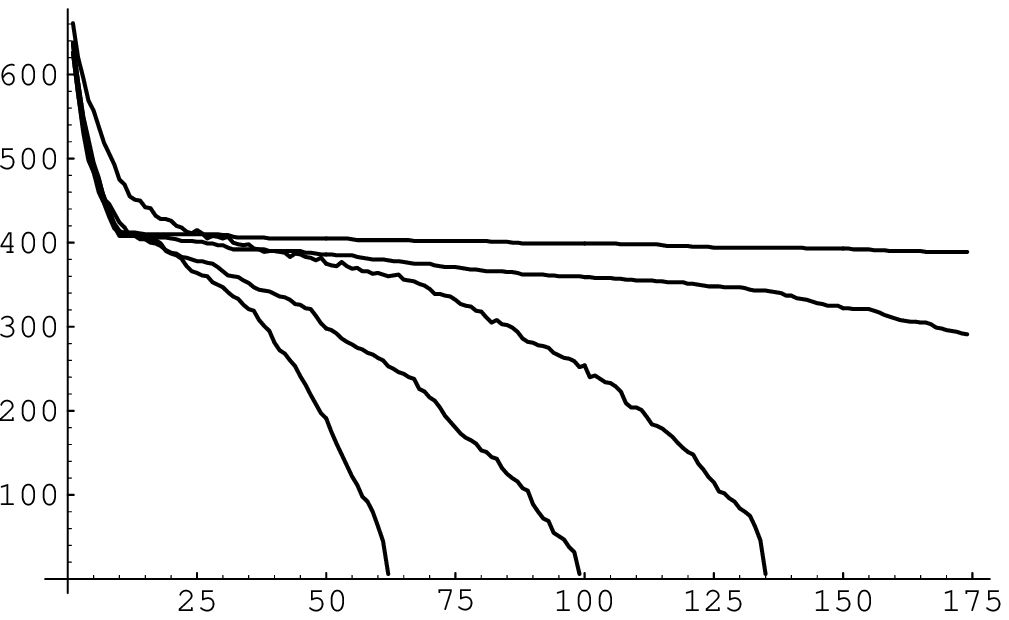}}}
\end{figure}
\noindent
{\bf Fig.4} \\
The same as figure 3, dependence $N_0$ {\em vs.} time.\\

\end{document}